\documentclass[a4paper,11pt]{article}
\usepackage{bm,amsmath,amssymb,mathrsfs,graphicx,citesort}
\newcommand{\nc}{\newcommand}
\nc{\rnc}{\renewcommand}
\nc{\nn}{\nonumber}
\nc{\bra}{\langle}
\nc{\ket}{\rangle}
\rnc{\(}{\left(}
\rnc{\)}{\right)}
\rnc{\[}{\left[}
\rnc{\]}{\right]}

\usepackage[dvips]{color}
\nc{\tcr}{\textcolor{red}}

\begin{document}%
%
\title{
Absence of finite size correction at the combinatorial point
of the integrable higher spin XXZ chain
}
\author{
Kohei Motegi\thanks{E-mail: motegi@gokutan.c.u-tokyo.ac.jp} \,
\\\\
Okayama Institute for Quantum Physics, \\
 Kyoyama 1-9-1, Okayama 700-0015, Japan \\
\\\\
\\
}

\date{\today}
 
 
\maketitle

%
%
\begin{abstract}
We investigate the integrable higher spin XXZ chain at the
Razumov-Stroganov point.
We present a method to evaluate the exact value
of the eigenvalue which is conjectured to correspond
to the groundstate of the Hamiltonian for finite size chain
from the Baxter $Q$ operator.
This allows us to examine the
exact total energy difference between different number of total sites,
from which we find strong evidence for the absence of finite size correction to
the groundstate energy.
\end{abstract}


\section{Introduction}
The spin 1/2 Heisenberg XXZ chain under the periodic boundary condition
\begin{align}
H(\eta)=-\frac{1}{2} \sum_{j=1}^M [ \sigma_j^x \sigma_{j+1}^x
+\sigma_j^y \sigma_{j+1}^y+ \mathrm{ch} \eta (\sigma_j^z \sigma_{j+1}^z-1)  ].
\end{align}
is the first discovered and is one of the most fundamental models in
quantum integrable systems.
The Hamiltonian can be diagonalized by the Bethe ansatz \cite{Bethe} to give
the eigenvalues
\begin{align}
\mathcal{E}(\eta)=\sum_{j=1}^M(2 \mathrm{ch} \eta -w_j-w_j^{-1}),
\end{align}
under the following constraints between the parameters $\{ w_j \}$
\begin{align}
w_j^M=\prod_{\substack{k=1 \\ k \neq j}}^p
\Bigg\{
- \frac{w_j w_k-2 \mathrm{ch} (\eta) w_j+1}{w_j w_k-2 \mathrm{ch} (\eta) w_k+1}
\Bigg\},
\end{align}
which is called the Bethe ansatz equation.
The groundstate energy can be calculated exactly in the thermodynamic limit
by making Fourier transform, but it is in general impossible
to find the exact value for finite size chain,
due to finite size correction.
However, 
taking the anisotropy parameter to a special value $\eta=-2 \pi i/3$,
the total number of sites to be odd $M=2N+1$ and the total spin
to be $S_z^{\mathrm{tot}}=1/2 \ (p=N)$,
the groundstate energy is shown to have no finite size correction 
\cite{St,RS,RS2}
\begin{align}
\mathcal{E}(-2 \pi i/3 )=-M.
\end{align}
For the last ten years,
many exact results such as
the components of the groundstate wavefunction and
the emptiness formation probability
were conjectured to be related to combinatorial objects
such as the alternating sign matrix 
\cite{RS,RS2,DBNM,BCV}, some of them have been proved 
\cite{DZZ,CS,C}.
Nowadays, the special point of
the anisotropy parameter $\eta=-2 \pi i/3$
is called the combinatorial point or the Razumov-Stroganov point.

In this paper, we describe a method to evaluate the
eigenvalue which is conjectured to correspond to the
groundstate of the Hamiltonian of
the integrable higher half-integer spin-$s$ XXZ chain of finite size at the
Razumov-Stroganov point. By appropriately normalizing the overall factor
of the Hamiltonian which does not depend on the length of the chain,
the eigenvalue of the Hamiltonian can be calculated as
\begin{align}
\mathcal{E}(\eta)=\sum_{j=1}^M(2 \mathrm{ch} (2 s \eta) -w_j-w_j^{-1}),
\label{eigenvalueexp}
\end{align}
under the Bethe ansatz equation
\begin{align}
w_j^M=\prod_{\substack{k=1 \\ k \neq j}}^p
\Bigg\{
- \frac{
\mathrm{sh} (\eta) w_j w_k- \mathrm{sh} ((2s+1) \eta) w_j
+\mathrm{sh} ((2s-1) \eta) w_k +\mathrm{sh} \eta
}
{
\mathrm{sh} (\eta) w_j w_k- \mathrm{sh} ((2s+1) \eta) w_k
+\mathrm{sh} ((2s-1) \eta) w_j +\mathrm{sh} \eta
}
\Bigg\}. \label{BAE}
\end{align}
As mentioned before, one cannnot evaluate the eigenvalue
exactly for finite size chain in general.
However, for odd number of sites and at the
Razumov-Stroganov point, we recently calculated a polynomial called
the $Q$ operator whose zeros give the Bethe roots \cite{Motegi}.
This was obtained by solving the Baxter $TQ$ equation \cite{Baxter,DST}
which is essentially equivalent to the Bethe ansatz equation.
The computed $Q$ operator is based on the variables
$z_j=(w_j-\mathrm{e}^{-2 s \eta})/(\mathrm{e}^{-2 s \eta}w_j-1)$
rather than the variables $w_j$.
To evaluate the exact eigenvalue of the Hamiltonian
from the obtained $Q$ operator, 
we switch from the $Q$ operator in the variables $z_j$
to the one in the variables $w_j$.
This leads to the transformation of the symmetric polynomials
of the variables $z_j$ to those of the variables $w_j$,
and the simplest symmetric polynomial $\sum_{j=1}^p w_j$
essentially leads to the eigenvalue of the Hamiltonian.
By this procedure, we can easily evaluate the groundstate energy exactly
for finite chain, and one finds strong evidence for the absence
of finite size correction for higher half-integer spin XXZ chain at the
Razumov-Stroganov point,
generalizing the behavior observed in the spin 1/2 XXZ chain.
Calculating the groundstate eigenvalue naively by solving the
zeros of the $Q$ operator just only gives a numerical value.
The transformation of the $Q$ operators is essential to extract
the exact value, from which one can check the validity
of the vanishing of finite size correction.
Based on the conjecture for the absence of finite size correction, 
it is enough to consider only $M=3$ sites $(N=1)$
and $M=5$ sites $(N=2)$ to extract the groundstate energy
which can be easily evaluated by the $Q$ operator.

\section{$TQ$ equation and $Q$ operator}
In this section,
we give a brief review on the results for the $Q$ operator
of higher spin XXZ chain at the Razumov-Stroganov point \cite{Motegi}.
In terms of the variables 
$z_j=(w_j-\mathrm{e}^{-2 s \eta})/(\mathrm{e}^{-2 s \eta}w_j-1)$,
the Bethe ansatz equation \eqref{BAE} can be expressed as
\begin{align}
\Bigg( \frac{z_j \mathrm{e}^{2s \eta}-1}{z_j-\mathrm{e}^{2s \eta}} \Bigg)^M
=\prod_{\substack{k=1 \\ k \neq j}}^p
\Bigg(
\frac{z_j \mathrm{e}^{2 \eta}-z_k}{z_j-z_k \mathrm{e}^{2 \eta}}
\Bigg),
\end{align}
or, in the familiar form
\begin{align}
\Bigg( \frac{\mathrm{sh}(u_j+s \eta)}{\mathrm{sh}(u_j-s \eta)} \Bigg)^M
=\prod_{\substack{k=1 \\  k \neq j}}^p
\frac{\mathrm{sh}(u_j-u_k+\eta)}{\mathrm{sh}(u_j-u_k-\eta)},
\end{align}
in terms of the variables $u_j$ related to $z_j$ by $z_j=\mathrm{exp}(2u_j)$.
\\
The Bethe ansatz equation can be obtained from the
Baxter's $TQ$ equation
\begin{align}
T(u)Q(u)=\mathrm{sh}^{M}(u+s \eta) Q(u-\eta)
+\mathrm{sh}^{M}(u-s \eta) Q(u+\eta)=0,
\label{TQ}
\end{align}
where  $T(u)$ is the transfer matrix 
whose auxiliary space has spin $1/2$ and
the quantum space is the $M$-fold tensor product of spin $s$ spaces.
$u$ is the spectral parameter and
$\eta$ is the anisotropy parameter associated with the XXZ chain.
The $Q$ operator
$Q(u)=\prod_{j=1}^p \mathrm{sh}(u-u_j)$ encodes the information of the
eigenstate of the model since
the $TQ$ equation \eqref{TQ} reduces to the Bethe ansatz equation
of the higher spin XXZ chain by setting the spectral parameter $u$ to $u=u_j$.
Solving the Bethe ansatz equation
is equivalent to computing the $Q$ operator.

For the half-integer spin $s=(L-2)/2 \ (L=3,5,7, \cdots)$ XXZ chain
with odd number of total sites $M=2N+1 \ (N=1,2,3, \cdots)$,
the transfer matrix eigenvalue of  $T(u)$ which is conjectured to
correspond to the groundstate was found to have
a simple form at the Razumov-Stroganov point $\eta=-(L-1) \pi i/L$.
In the sector with $p=N+(2N+1)(L-3)/2+m$ Bethe roots,
an exact transfer matrix eigenvalue has the following simple form \cite{DST}
\begin{align}
T(u)=2 \mathrm{ch} \Bigg( \frac{(L-1)(1-2m) \pi i}{2L} \Bigg) \mathrm{sh}^M u.
\label{transfereigenvalue}
\end{align}
To analyze the corresponding $Q$ operator,
it is useful to change to the spectral variables
$z=\mathrm{exp}(2u)$ and $z_j=\mathrm{exp}(2u_j)$ and
redefine the $Q$ operator as
\begin{align}
Q(z)&=\prod_{j=1}^p(z-z_j)=\sum_{j=0}^p (-1)^j z^{p-j} e_j^{(p)}, \\
e_j^{(p)}&=\sum_{1 \le i_1 < i_2 < \cdots < i_j \le p}
z_{i_1} z_{i_2} \cdots z_{i_j} \ (j=1,2,\cdots,p), \ e_0^{(p)}=1.
\end{align}
We recently calculated the $Q$ operator in two ways.
Let us briefly present the results below.
We restrict to the case $m=0$, i.e.,
in the sector $S_z^\mathrm{tot}=1/2 \ (p=N+(2N+1)(L-3)/2)$.
First, we showed that solving the $TQ$ equation reduces to
solving the following set of linear equations of the
elementary symmetric polynomials $e_k^{(p)}$ of $z_j$
\begin{align}
\sum_{j=\mathrm{max}(0,\ell-2N-1)}^{\mathrm{min}(N+(2N+1)(L-3)/2,\ell)}
\binom{2N+1}{\ell-j} e_j^{(N+(2N+1)(L-3)/2)}=0, \label{rewriteTQ2}
\end{align}
for 
$\ell=0,1,\cdots,NL+(L-1)/2 \ (\ell \neq Lk,Lk+(L-1)/2 \ (k=0,1,\cdots,N))$. \\
We also evaluated the $Q$ operator in another way by use of the
interpolation formula to find
\begin{align}
Q(z)=&(z-1)^{-(2N+1)}
\Bigg\{
\sum_{k=0}^{N/2}(-1)^k \binom{N}{k} \prod_{j=0}^N 
\frac{(L-1)/2+Lj}{(L-1)/2-Lk+Lj}
(z^{LN+(L-1)/2-Lk}-z^{Lk}) \nonumber \\
&+
\sum_{k=0}^{N/2-1}(-1)^k \binom{N}{k} \prod_{j=0}^N 
\frac{(L-1)/2+Lj}{-(L-1)/2-Lk+Lj}
(z^{LN-Lk}-z^{Lk+(L-1)/2})
\Bigg\}, \label{Qoperatoreven}
\end{align}
for $N$ even and
\begin{align}
Q(z)=&(z-1)^{-(2N+1)}
\Bigg\{
\sum_{k=0}^{(N-1)/2}(-1)^k \binom{N}{k} \prod_{j=0}^N 
\frac{(L-1)/2+Lj}{(L-1)/2-Lk+Lj}
(z^{LN+(L-1)/2-Lk}-z^{Lk}) \nonumber \\
&+
\sum_{k=0}^{(N-1)/2}(-1)^k \binom{N}{k} \prod_{j=0}^N 
\frac{(L-1)/2+Lj}{-(L-1)/2-Lk+Lj}
(z^{LN-Lk}-z^{Lk+(L-1)/2})
\Bigg\}, \label{Qoperatorodd}
\end{align}
for $N$ odd.
\section{Transformation of $Q$ operators}
It was useful to change to the variables
$z$ and $z_j$ to calculate the $Q$ operator.
On the other hand, it is essential to use the variables
$w=(z \mathrm{e}^{2s \eta}-1)/(z-\mathrm{e}^{2s \eta})$
and $w_j=(z_j \mathrm{e}^{2s \eta}-1)/(z_j-\mathrm{e}^{2s \eta})$
to evaluate the exact eigenvalue of the Hamiltonian \cite{RS2}.
This is because $E_1^{(p)}=\sum_{j=1}^p w_j$
which forms a part of the eigenvalue \eqref{eigenvalueexp}
is the coefficient of $w^{p-1}$ of the following
$Q$ operator in the $w$ variables
\begin{align}
\chi(w)&=\prod_{j=1}^p (w-w_j)=
\sum_{j=0}^p (-1)^j w^{p-j} E_j^{(p)}, \label{qoperatorw} \\
E_j^{(p)}&=\sum_{1 \le i_1 < i_2 < \cdots < i_j \le p}
w_{i_1} w_{i_2} \cdots w_{i_j} \ (j=1,2,\cdots,p), \ E_0^{(p)}=1.
\end{align}
To evaluate the exact eigenvalue,
we need to express $E_1^{(p)}$ in terms of the symmetric polynomials 
$e_k^{(p)}$ of the variables $z_j$.
We insert the relation
\begin{align}
w-w_j=\frac{(\mathrm{e}^{-4 \pi i/L}-1)(z-z_j)}{(z-\mathrm{e}^{-2 \pi i/L})(\mathrm{e}^{-2 \pi i/L}-z_j)},
\end{align}
into \eqref{qoperatorw} and make the following transformation
to express the coefficients of the powers of $w$ in terms of
$e_k^{(p)}$
\begin{align}
\chi(w)&=
\Bigg(
\frac{\mathrm{e}^{-4 \pi i/L}-1}{z-\mathrm{e}^{-2 \pi i/L}}
\Bigg)^p \frac{Q(z)}{Q(\mathrm{e}^{-2 \pi i/L})} \nonumber \\
&=
(w-\mathrm{e}^{-2 \pi i/L})^p
\frac{Q \Big( \frac{\mathrm{e}^{- 2 \pi i/L}w-1}
{w-\mathrm{e}^{-2 \pi i/L}} \Big)}{Q(\mathrm{e}^{-2 \pi i/L})} \nonumber \\
&=\frac{(w-\mathrm{e}^{-2 \pi i/L})^p}{Q(\mathrm{e}^{-2 \pi i/L})}
\prod_{j=1}^p \Bigg( 
\frac{\mathrm{e}^{- 2 \pi i/L}w-1}
{w-\mathrm{e}^{-2 \pi i/L}}
-z_j \Bigg) \nonumber \\
&=\frac{1}{Q(\mathrm{e}^{-2 \pi i/L})}
\prod_{j=1}^p
(\mathrm{e}^{-2 \pi i/L}w-1
-z_j(w-\mathrm{e}^{-2 \pi i/L})
) \nonumber \\
&=\frac{1}{Q(\mathrm{e}^{-2 \pi i/L})}
\sum_{k=0}^p
(\mathrm{e}^{-2 \pi i/L}w-1)^{p-k}
(\mathrm{e}^{-2 \pi i/L}-w)^k e_k^{(p)} \nonumber \\
&=\frac{1}{Q(\mathrm{e}^{-2 \pi i/L})}
\sum_{k=0}^p \sum_{\ell=0}^{p-k} \sum_{j=0}^k
(-1)^{p+j-k-\ell}
\mathrm{e}^{-2 \pi i(k+\ell-j)/L}
\binom{p-k}{\ell} \binom{k}{j} e_k^{(p)} w^{j+\ell} \nonumber \\
&=\frac{1}{Q(\mathrm{e}^{-2 \pi i/L})}
\sum_{k=0}^p \sum_{\alpha=0}^p 
\sum_{j=\mathrm{max}(0,k-\alpha)}^{\mathrm{min}(k,p-\alpha)}
(-1)^{\alpha+k}
\mathrm{e}^{-2 \pi i(k+p-\alpha-2 j)/L}
\binom{p-k}{p-\alpha-j} \binom{k}{j} w^{p-\alpha} e_k^{(p)} \label{wexpansion}.
\end{align}
Equating the coefficients of the powers of $w$ of
\eqref{qoperatorw} and \eqref{wexpansion}, one gets
\begin{align}
E_\alpha^{(p)}=\frac{1}{Q(\mathrm{e}^{-2 \pi i/L})}
\sum_{k=0}^p \sum_{j=\mathrm{max}(0,k-\alpha)}^{\mathrm{min}(k,p-\alpha)}
(-1)^k
\mathrm{e}^{-2 \pi i(k+p-\alpha-2 j)/L}
\binom{p-k}{p-\alpha-j} \binom{k}{j} w^{p-\alpha} e_k^{(p)}.
\end{align}
In particular, we have
\begin{align}
E_1^{(p)}=\frac{1}{Q(\mathrm{e}^{-2 \pi i/L})}
\sum_{k=0}^p \sum_{j=\mathrm{max}(0,k-1)}^{\mathrm{min}(k,p-1)}
(-1)^k
\mathrm{e}^{-2 \pi i(k+p-1-2 j)/L}
\binom{p-k}{p-1-j} \binom{k}{j} w^{p-1} e_k^{(p)}.
\end{align}
We can furthermore simplify the numerator and the denominator utilizing the
relation 
\begin{align}
e_k^{(p)}=(-1)^p e_{p-k}^{(p)}.
\end{align}
This relation can be shown by
combining
\begin{align}
e_k^{(p)}|_{z_j \to z_j^{-1}}&=
\sum_{1 \le i_1 < i_2 < \cdots < i_k \le p}
z_{i_1}^{-1} z_{i_2}^{-1} \cdots z_{i_k}^{-1} \nonumber \\
&=\sum_{1 \le i_1 < i_2 < \cdots < i_k \le p}
z_{i_1} z_{i_2} \cdots z_{i_k}
=e_k^{(p)}, \label{symminv}
\end{align}
which follows from the $z \leftrightarrow z^{-1}, z_j \leftrightarrow z_j^{-1}$
invariance of the $TQ$ equation
\begin{align}
&-2 \mathrm{ch} \Bigg( \frac{(L-1) \pi i}{2L} \Bigg)(z-1)^{2N+1}
\prod_{j=1}^{N+(2N+1)(L-3)/2}(z-z_j) \nonumber \\
+&
\mathrm{e}^{(1-L)\pi i/2L}(z-\mathrm{e}^{2 \pi i/L})^{2N+1}
\prod_{j=1}^{N+(2N+1)(L-3)/2}(z-\mathrm{e}^{2 \pi i/L} z_j)
\nonumber \\
+&\mathrm{e}^{(L-1)\pi i/2L}(z-\mathrm{e}^{-2 \pi i/L})^{2N+1}
\prod_{j=1}^{N+(2N+1)(L-3)/2}(z-\mathrm{e}^{-2 \pi i/L} z_j)=0,
\end{align}
and
\begin{align}
Q(0)=(-1)^p \prod_{j=1}^p z_j=1,
\end{align}
which follows by substituting $z=0$ in 
\eqref{Qoperatoreven} and \eqref{Qoperatorodd}, as
\begin{align}
e_k^{(p)}&=\sum_{1 \le i_1 < i_2 < \cdots < i_k \le p}
z_{i_1} z_{i_2} \cdots z_{i_k} \nonumber \\
&=\prod_{j=1}^p z_j
\sum_{1 \le i_1 < i_2 < \cdots < i_{p-k} \le p}
z_{i_1}^{-1} z_{i_2}^{-1} \cdots z_{i_{p-k}}^{-1} \nonumber \\
&=(-1)^p
\sum_{1 \le i_1 < i_2 < \cdots < i_{p-k} \le p}
z_{i_1} z_{i_2} \cdots z_{i_{p-k}} \nonumber \\
&=(-1)^p e_{p-k}^{(p)}.
\end{align}
Using this relation,
the numerator can be expressed as
\begin{align}
&\sum_{k=0}^p \sum_{
j=\mathrm{max}(0,k-1)}^{\mathrm{min}(k,p-1)}
(-1)^k
\mathrm{e}^{-2 \pi i(k+p-1-2 j)/L}
\binom{p-k}{p-1-j} \binom{k}{j} w^{p-1} e_k^{(p)} \nonumber \\
=&2 \mathrm{e}^{-\pi p i/L}
\sum_{k=0}^{p-1} (-1)^k (p-k) 
\mathrm{cos} \Bigg( \frac{\pi (2k+2-p)}{L} \Bigg) e_k^{(p)},
\label{num}
\end{align}
and the denominator as
\begin{align}
Q(\mathrm{e}^{-2 \pi i/L})
=\mathrm{e}^{-\pi p i/L}
\sum_{k=0}^p (-1)^k 
\mathrm{cos} \Bigg( \frac{\pi(p-2k)}{L} \Bigg) e_k^{(p)}.
\label{den}
\end{align}
The denominator can also be evaluated by setting
$z=\mathrm{e}^{-2 \pi i/L}$ in
\eqref{Qoperatoreven} or \eqref{Qoperatorodd}.
The factor $\mathrm{e}^{-\pi p i/L}$
in \eqref{num} and \eqref{den} cancels out and we have
\begin{align}
E_1^{(p)}=\sum_{j=1}^p w_j=\frac{
2 \sum_{k=0}^{p-1} (-1)^k (p-k) 
\mathrm{cos} \Big( \frac{\pi (2k+2-p)}{L} \Big) e_k^{(p)}
}
{
\sum_{k=0}^p (-1)^k 
\mathrm{cos} \Big( \frac{\pi(p-2k)}{L} \Big) e_k^{(p)}
}. \label{transform}
\end{align}
This is the expression of the term $E_1^{(p)}=\sum_{j=1}^p w_j$
in terms of the symmetric polynomials $e_k^{(p)}$
of the variables $z_k$.
The other term $\sum_{j=1}^p w_j^{-1}$ which consists
another part of the eigenvalue of the Hamiltonian \eqref{eigenvalueexp} gives the same value
with $E_1^{(p)}$
\begin{align}
\sum_{j=1}^p w_j=\sum_{j=1}^p w_j^{-1}.
\end{align}
This can be shown by utilizing the invariance \eqref{symminv}
and noting that \eqref{transform} replaced by
$w_j \to w_j^{-1}$ and $z_j \to z_j^{-1}$
holds by comparing
$w_j=(z_j \mathrm{e}^{2 s \eta}-1)/(z_j-\mathrm{e}^{2 s \eta})$ and
$w_j^{-1}=(z_j^{-1} \mathrm{e}^{2 s \eta}-1)/(z_j^{-1}-\mathrm{e}^{2 s \eta})$
\begin{align}
\sum_{j=1}^p w_j^{-1}&=\frac{
2 \sum_{k=0}^{p-1} (-1)^k (p-k) 
\mathrm{cos} \Big( \frac{\pi (2k+2-p)}{L} \Big) 
e_k^{(p)}|_{z_j \to z_j^{-1}}
}
{
\sum_{k=0}^p (-1)^k 
\mathrm{cos} \Big( \frac{\pi(p-2k)}{L} \Big) 
e_k^{(p)}|_{z_j \to z_j^{-1}}
} \nonumber \\
&=\frac{
2 \sum_{k=0}^{p-1} (-1)^k (p-k) 
\mathrm{cos} \Big( \frac{\pi (2k+2-p)}{L} \Big) 
e_k^{(p)}
}
{
\sum_{k=0}^p (-1)^k 
\mathrm{cos} \Big( \frac{\pi(p-2k)}{L} \Big) 
e_k^{(p)}
} \nonumber \\
&=\sum_{j=1}^p w_j.
\end{align}
\section{Absence of finite size correction}
The procedure to calculate the exact eigenvalue of the Hamiltonian
for finite size chain
can be summarized
as: \\
\\
(i)Solve the $TQ$ equation to calculate the $Q$ operator
in the $z$ variables.
This reduces to solving the linear equations 
\eqref{rewriteTQ2}
of the symmetric polynomials
$e_j^{(p)}$ of the $z$ variables, or to
solve the $TQ$ equation explicitly 
by use of the interpolation formula to get
\eqref{Qoperatoreven} or \eqref{Qoperatorodd}. \\
\\
(ii)Relate the $Q$ operator in the $z$ variables and the $w$ variables
and express the symmetric polynomials $E_j^{(p)}$ of the $w$ variables
in terms of the symmetric polynomials $e_k^{(p)}$ of the $z$ variables.
\\
\\
(iii)Evaluate the exact eigenvalue \eqref{eigenvalueexp}
by inserting the exact value of $e_k^{(p)}$ computed in (i)
to the expression \eqref{transform}
relating $E_{1}^{(p)}$ and $e_k^{(p)}$ which is done in (ii). \\
\\
For spin 1/2 $(L=3)$, it is shown \cite{RS2} that
\begin{align}
\sum_{j=1}^p w_j=\frac{1}{2}+\frac{N}{2}, \label{sum}
\end{align}
by identifying the symmetric polynomials
in the $w$ variables with the refined enumuration
of the alternating sign matrix.
\eqref{sum} can also be observed by the procedure described above.
This leads to the absence of finite size correction
to the groundstate energy at the Razumov-Stroganov point
\begin{align}
\mathcal{E}(-2 \pi i/3)=-M.
\end{align}

Next, we examine the spin 3/2 $(L=5)$ chain. 
By calculating $E_1^{(p)}$ for various
number of total sites,
we conjecture that the following relation
\begin{align}
\sum_{j=1}^p w_j=\frac{1+\sqrt{5}}{2}+\frac{3+5 \sqrt{5}}{4}N,
\end{align}
holds.
This leads to the following conjecture for the eigenvalue
\begin{align}
\mathcal{E}(-4 \pi i/5)=-\frac{3+\sqrt{5}}{2}M,
\end{align}
which means there is no finite size correction to the groundstate energy
for the spin 3/2 chain
at the Razumov-Stroganov point $\eta=-4 \pi i/5$ as well.

By examining larger spins, we conjecture that
$E_1^{(p)}$ can be expressed as
\begin{align}
\sum_{j=1}^p w_j =A+\Bigg\{ 2A+\mathrm{cos} 
\Bigg( \frac{2 \pi}{L} \Bigg) \Bigg\} N,
\end{align}
where $A$ depends only on the spin value.
One can see $A=1/2$ for spin 1/2
and $A=(1+\sqrt{5})/2$ for spin 3/2 as above.
We have checked this for various spins and total number of sites.
This leads to the absence of finite size correction
to the groundstate energy
\begin{align}
\mathcal{E}(-(L-1) \pi i/L)=\Bigg((L-3) \mathrm{cos}
\Bigg( \frac{2 \pi}{L} \Bigg)
-2A \Bigg)M. \label{energyfinitesize}
\end{align}
Based on the conjecture for the absence of
finite size correction, $E_1^{(p)}=\sum_{j=1}^p w_j$
and the groundstate energy can be evaluated by only
treating $M=3$ sites $(N=1)$ and $M=5$ sites $(N=2)$.
Here is a list of $\sum_{j=1}^p w_j$ for spin 5/2 $(L=7)$, spin 7/2 $(L=9)$
and spin 9/2 $(L=11)$.
Note that the value $A$ in \eqref{energyfinitesize} is obtained by substituting
$N=0$ in the expression for $\sum_{j=1}^p w_j$ listed below.
\\
\\
spin 5/2 $(L=7)$
\begin{align}
\sum_{j=1}^p w_j
=&(N-1) \frac{6(-499+525 \mathrm{cos}(\pi/7)+
694 \mathrm{sin}(\pi/14)-900 \mathrm{sin}(3 \pi/14))}
{-235+290 \mathrm{cos}(\pi/7)
+350 \mathrm{sin}(\pi/14)-434 \mathrm{sin}(3 \pi/14)}
\nonumber \\
&+(2-N) \frac{-6+120 \mathrm{cos}(\pi/7)
+81 \mathrm{sin}(\pi/14)-38 \mathrm{sin}(3 \pi/14)}
{
-2+15 \mathrm{cos}(\pi/7)+12 \mathrm{sin}(\pi/14)-6 \mathrm{sin}(3 \pi/14)
}.
\end{align}
spin 7/2 $(L=9)$
\begin{align}
\sum_{j=1}^p w_j
=&(N-1) 
\frac{7695+43820 \mathrm{cos}(\pi/9)
-26108 \mathrm{cos}(2 \pi/9)-32210 \mathrm{sin}(\pi/18)}
{
351+2450 \mathrm{cos}(\pi/9)-1640 \mathrm{cos}(2 \pi/9)
-1910 \mathrm{sin}(\pi/18)
} \nonumber \\
&+(2-N)
\frac{-459+250 \mathrm{cos}(\pi/9)
-1360 \mathrm{cos}(2 \pi/9)-976 \mathrm{sin}(\pi/18)}
{
-36+40 \mathrm{cos}(\pi/9)-124 \mathrm{cos}(2 \pi/9)
-100 \mathrm{sin}(\pi/18)
}.
\end{align}
spin 9/2 $(L=11)$
\begin{align}
&\sum_{j=1}^p w_j
=(N-1) 
\frac{B}{C}+(2-N) \nonumber \\
\times&\frac{-75+860 \mathrm{cos}(\pi/11)
-207 \mathrm{cos}(2 \pi/11)+575 \mathrm{sin}(\pi/22)
-378 \mathrm{sin}(3 \pi/22)+765 \mathrm{sin}(5 \pi/22)}
{
-9+60 \mathrm{cos}(\pi/11)-21 \mathrm{cos}(2 \pi/11)
+45 \mathrm{sin}(\pi/22)-30 \mathrm{sin}(3 \pi/22)
+55 \mathrm{sin}(5 \pi/22)
}, \nonumber \\
B&=
-8675+9780 \mathrm{cos}(\pi/11)-16727 \mathrm{cos}(2 \pi/11)
+12895 \mathrm{sin}(\pi/22) \nonumber \\
&-15050 \mathrm{sin}(3 \pi/22)+10925 \mathrm{sin}(5 \pi/22)
\nonumber \\
C&=-376+480 \mathrm{cos}(\pi/11)-730 \mathrm{cos}(2 \pi/11)
+590 \mathrm{sin}(\pi/22) \nonumber \\
&-670 \mathrm{sin}(3 \pi/22)+520 \mathrm{sin}(5 \pi/22).
\end{align}

\section{Conclusion}
In this paper, we described a method to evaluate the
groundstate eigenvalue of the integrable higher spin XXZ chain
at the Razumov-Stroganov point by the $Q$ operator.
Since the spectral variables of the Bethe roots
convenient to evaluate the $Q$ operator and
the Hamiltonian are different, we made transformation
of the $Q$ operators of the two types of spectral variables.
This enabled us to get the exact groundstate eigenvalue,
which gives strong evidence for the absence of finite
size correction for higher half-integer integrable spin chain,
generalizing the behavior observed in the spin 1/2 chain.
Calculating the groundstate eigenvalue naively by solving the
zeros of the $Q$ operator just only gives a numerical value,
and the transformation of the $Q$ operators is essential to extract
the exact value.
For the spin 1/2 chain, this lead to the identification of the
symmetric polynomials of the Bethe roots to
the refined enumuration of alternating sign matrix \cite{RS2}.
It may be possible to prove for higher spins by considering
alternating sign matrix for higher spins.
Alternating sign matrix for spin 1 has been considered in \cite{BK}.
Another promising approach is the investigation
from the supersymmetric point of view \cite{FNS,YF}.
The eigenstate absent from finite size correction should
correspond to the zero energy groundstate of a
Hamiltonian which have supersymmetry.
For higher spins, progresses have been made in \cite{FNS,H}.

\section*{Acknowledgment}
The author thanks H. Katsura, A. Kuniba and K. Sakai
for useful discussions.


\end{document}